\documentclass[runningheads]{llncs}
\usepackage{graphicx}
\usepackage{textcomp}
\usepackage{siunitx}
\usepackage{amsmath}
\usepackage{multirow}
\usepackage{booktabs}

\begin{document}
\title{T2 Mapping from Super-Resolution-Reconstructed Clinical Fast Spin Echo Magnetic Resonance Acquisitions}
\titlerunning{Super-Resolution T2 Mapping}
%
\author{H\'el\`ene Lajous\inst{1,2} \and
Tom Hilbert\inst{1,3,4} \and
Christopher W. Roy\inst{1} \and
S\'ebastien Tourbier\inst{1} \and
Priscille de Dumast\inst{1,2} \and
Thomas Yu\inst{4} \and
Jean-Philippe Thiran\inst{1,4} \and
Jean-Baptiste Ledoux\inst{1,2} \and
Davide Piccini\inst{1,3} \and
Patric Hagmann\inst{1} \and
Reto Meuli\inst{1} \and
Tobias Kober\inst{1,3,4} \and
Matthias Stuber\inst{1,2} \and
Ruud B. van Heeswijk\inst{1} \and
Meritxell Bach Cuadra\inst{1,2,4}}
%
%
\authorrunning{H. Lajous et al.}
%
\institute{Department of Radiology, Lausanne University Hospital (CHUV) and University of Lausanne (UNIL), Lausanne, Switzerland\\
\email{helene.lajous@unil.ch} \and
Center for Biomedical Imaging (CIBM), Lausanne, Switzerland \and
Advanced Clinical Imaging Technology (ACIT), Siemens Healthcare, Lausanne, Switzerland \and
Signal Processing Laboratory 5 (LTS5), Ecole Polytechnique F\'ed\'erale de Lausanne (EPFL), Lausanne, Switzerland}
\maketitle              
\begin{abstract}
Relaxometry studies in preterm and at-term newborns have provided insight into brain microstructure, thus opening new avenues for studying normal brain development and supporting diagnosis in equivocal neurological situations. However, such quantitative techniques require long acquisition times and therefore cannot be straightforwardly translated to \emph{in utero} brain developmental studies. In clinical fetal brain magnetic resonance imaging routine, 2D low-resolution T2-weighted fast spin echo sequences are used to minimize the effects of unpredictable fetal motion during acquisition. As super-resolution techniques make it possible to reconstruct a 3D high-resolution volume of the fetal brain from clinical low-resolution images, their combination with quantitative acquisition schemes could provide fast and accurate T2 measurements. In this context, the present work demonstrates the feasibility of using super-resolution reconstruction from conventional T2-weighted fast spin echo sequences for 3D isotropic T2 mapping. A quantitative magnetic resonance phantom was imaged using a clinical T2-weighted fast spin echo sequence at variable echo time to allow for super-resolution reconstruction at every echo time and subsequent T2 mapping of samples whose relaxometric properties are close to those of fetal brain tissue. We demonstrate that this approach is highly repeatable, accurate and robust when using six echo times (total acquisition time under 9 minutes) as compared to gold-standard single-echo spin echo sequences (several hours for one single 2D slice).
\end{abstract}
\section{Introduction}
Early brain development encompasses many crucial structural and physiological modifications that have an influence on health later in life. Changes in T1 and T2 relaxation times may provide valuable clinical information about ongoing biological processes, as well as a better insight into the early stages of normal maturation~\cite{deoni_quantitative_2010}. Indeed, quantitative MRI (qMRI) has revealed biomarkers sensitive to subtle changes in brain microstructure that are characteristic of abnormal patterns and developmental schemes in newborns~\cite{dingwall_t2_2016,schneider_evolution_2016}. T1 and T2 mapping of the developing fetal brain would afford physicians new resources for pregnancy monitoring, including quantitative diagnostic support in equivocal situations and prenatal counselling, as well as postnatal management. Unfortunately, current relaxometry strategies require long scanning times that are not feasible in the context of \emph{in utero} fetal brain MRI due to unpredictable fetal motion in the womb~\cite{chen_t2_2018,gholipour_fetal_2014,leppert_t2_2009,travis_more_2019}. As such, very little work has explored \emph{in vivo} qMRI of the developing fetal brain. Myelination was characterized \emph{in utero} using a mono-point T1 mapping based on fast spoiled gradient echo acquisitions ~\cite{abd_almajeed_myelin_2004}, and more recently by fast macromolecular proton fraction mapping~\cite{yarnykh_quantitative_2018}. T2* relaxometry of the fetal brain has been explored through fast single-shot multi-echo gradient echo-type echo-planar imaging (GRE-EPI)~\cite{vasylechko_t2_2015} and, recently, based on a slice-to-volume registration of 2D dual-echo multi-slice EPI with multiple time points reconstructed into a motion-free isotropic high-resolution (HR) volume~\cite{blazejewska_3d_2017}. To our knowledge, similar strategies have not been investigated for \emph{in utero} T2 mapping yet.
Today, super-resolution (SR) techniques have been adopted to take advantage of the redundancy between multiple T2-weighted (T2w) low-resolution (LR) series acquired in orthogonal orientations and thereby reconstruct a single isotropic HR volume of the fetal brain with reduced motion sensitivity for thorough anatomical exploration~\cite{ebner_automated_2020,gholipour_robust_2010,kainz_fast_2015,rousseau_super-resolution_2010,tourbier_efficient_2015}. In clinical routine, 2D thick slices are typically acquired in a few seconds using T2w multi-slice single-shot fast spin echo sequences~\cite{gholipour_fetal_2014}. We hypothesize that the combination of SR fetal brain MRI with the sensitivity of qMRI would enable reliable and robust 3D HR T2 relaxometry of the fetal brain~\cite{bano_model-based_2020,blazejewska_3d_2017}. In this context, we have explored the feasibility of repeatable, accurate and robust 3D HR T2 mapping from SR-reconstructed clinical fast T2w Half-Fourier Acquisition Single-shot Turbo spin Echo (HASTE) with variable echo time (TE) on a quantitative MR phantom~\cite{keenan_kathryn_e_multi-site_2016}.

\section{Methodology}
\subsection{Model Fitting for T2 Mapping}
\begin{figure}[hbt!]
\centering
\includegraphics[width=0.8\textwidth]{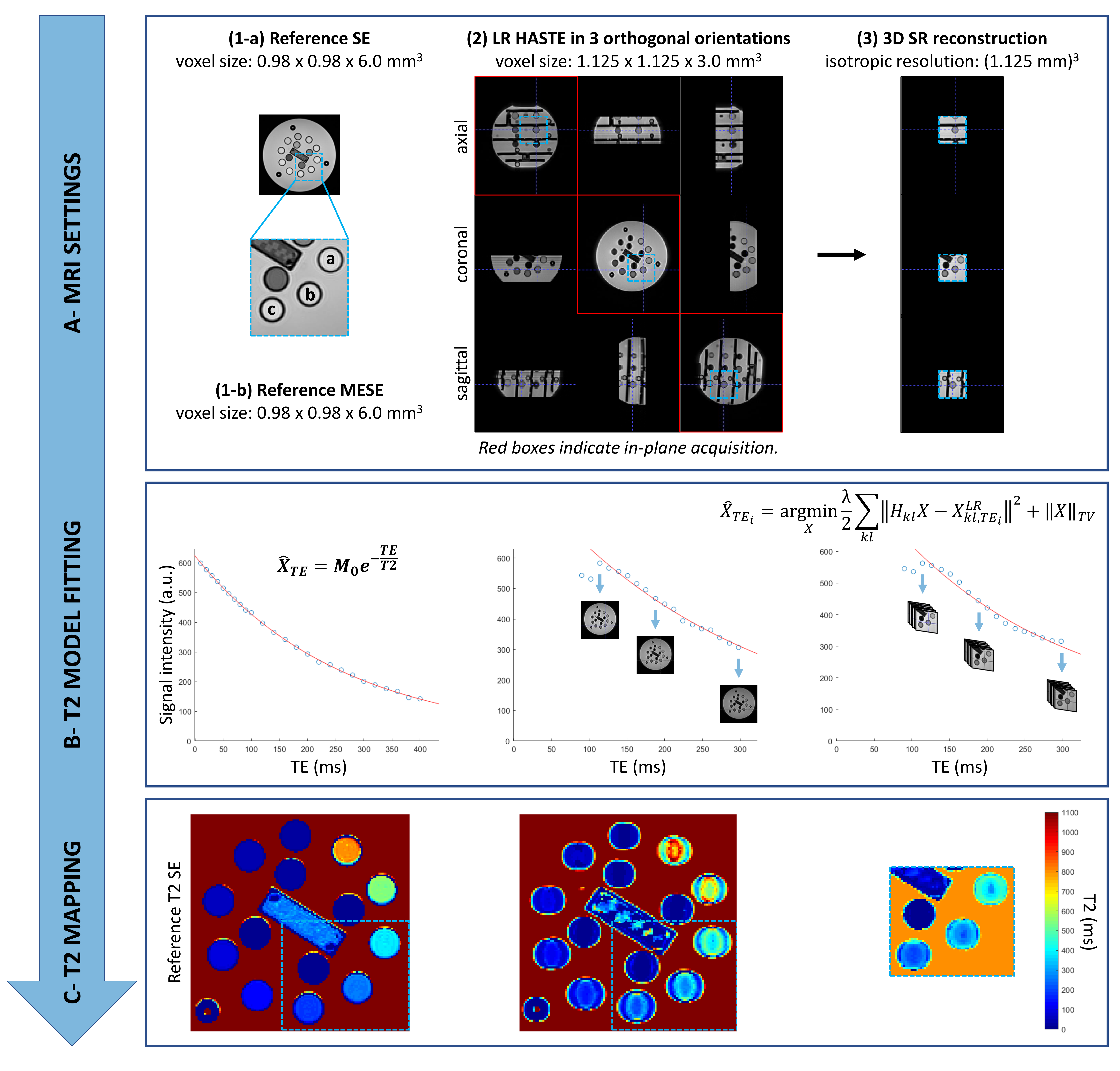}
\caption{Evaluation framework. Reference T2 values of elements (a), (b) and (c) (blue dashed area) of the NIST phantom are measured by (A-1-a) single-echo spin echo (SE) and (A-1-b) multi-echo spin echo (MESE) sequences. (A-2) Low-resolution orthogonal HASTE images acquired at variable TE are SR-reconstructed into (A-3) an isotropic volume for every TE. (B) The signal decay as a function of TE is fitted in each voxel by a mono-exponential model. (C) Resulting voxel-wise T2 maps.} \label{fig1}
\end{figure}

The T2w contrast of an MR image is governed by an exponential signal decay characterized by the tissue-specific relaxation time, T2. Since any voxel within brain tissue may contain multiple components, a multi-exponential model is the closest to reality. However, it requires long acquisition times that are not acceptable in a fetal imaging context. The common simplification of a single-compartment model~\cite{dingwall_t2_2016,leppert_t2_2009} allows for fitting the signal according to the following equation:
\begin{equation}
\hat{X}\textsubscript{TE} = \mathcal{M}\textsubscript{0} e^{\frac{-TE}{T2}},
\label{eq:model}
\end{equation}
where $\mathcal{M}$\textsubscript{0} is the equilibrium magnetization and $\hat{X}$\textsubscript{TE} is the signal intensity at a given echo time TE at which the image is acquired. As illustrated in Figure~\ref{fig1}, the time constant T2 can be estimated in every voxel by fitting the signal decay over TE with this mono-exponential analytical model~\cite{milford_mono-exponential_2015}.

We aim at estimating a HR 3D T2 map of the fetal brain with a prototype algorithm. Our strategy is based on SR reconstruction from orthogonal 2D multi-slice T2w clinical series acquired at variable TE (see complete framework in Figure~\ref{fig1}). For every $TE_{i}$, a motion-free 3D image $\mathbf{\hat{X}}_{TE_{i}}$ is reconstructed using a Total-Variation (TV) SR reconstruction algorithm~\cite{tourbier_sebastientourbiermialsuperresolutiontoolkit_2019,tourbier_efficient_2015} which solves:
\begin{equation}
\begin{aligned}
\mathbf{\hat{X}}_{TE_{i}} = \arg\min_{\mathbf{X}} \ \frac{\lambda}{2} \sum_{kl} \| \underbrace{\mathbf{D}_{kl}\mathbf{B}_{kl}\mathbf{M}_{kl}}_{\mathbf{H}_{kl}} \mathbf{X} - \mathbf{X}_{kl,TE_{i}}^{LR}\|^2 + \|\mathbf{X}\|_{TV},
\end{aligned}
\label{eq:sr}
\end{equation}
where the first term relates to data fidelity, $k$ being the $k$-th LR series $\mathbf{X}_{TE_{i}}^{LR}$ and $l$ the $l$-th slice. $\|\mathbf{X}\|_{TV}$ is a TV prior introduced to regularize the solution while $\lambda$ balances the trade-off between both data and regularization ($\lambda$=0.75). $\mathbf{D}$ and $\mathbf{B}$ are linear downsampling and Gaussian blurring operators given by the acquisition characteristics. $\mathbf{M}$, which encodes the rigid motion of slices, is set to the identity transform in the absence of motion.

The model fitting described in Equation~\ref{eq:model} is computed in every voxel of a SR 3D volume estimated at time TE. T2 maps are computed using a non-linear least-squares optimization (MATLAB, MathWorks, R2019a). As shown in Figure~\ref{fig1}-B, the T2 signal decay may reveal an offset between the first echoes and the rest of the curve that can be explained by stimulated echoes~\cite{mcphee_limitations_2018} and the sampling order of the HASTE sequence. It is common practice to exclude from the fitting the first points that exhibit the pure spin echo without the stimulated echo contributions~\cite{milford_mono-exponential_2015}.

\subsection{Validation Study}
\subsubsection{Quantitative Phantom.} Our validation is based on the system standard model 130 that was established by the National Institute for Standards and Technology (NIST) of the United States in collaboration with the International Society for Magnetic Resonance in Medicine (ISMRM). It is produced by QalibreMD (Boulder, CO, USA) and is hereafter referred to as the NIST phantom~\cite{keenan_kathryn_e_multi-site_2016}. This quantitative phantom was originally developed to assess the repeatability and reproducibility of MRI protocols across vendors and sites. Our study focuses on a region-of-interest (ROI) represented by a blue square in Figure~\ref{fig1}-A. It is centered on elements of the NIST phantom that have relaxometry properties close to those reported in the literature for \emph{in vivo} brain tissue of fetuses and preterm newborns at 1.5 T~\cite{blazejewska_3d_2017,hagmann_t2_2009,nossin-manor_quantitative_2013,vasylechko_t2_2015,yarnykh_quantitative_2018}, namely T2 values higher than 170 ms and 230 ms in grey matter and white matter respectively, and high T1 values. Accordingly, we focus on the following areas: (a) T2=428.3 ms, (b) T2=258.4 ms and (c) T2=186.1 ms, with a relatively high T1/T2 ratio (4.5-6.9), and which fall within a field-of-view similar to that of fetal MRI.

\subsubsection{MR Imaging.} Acquisitions are performed at 1.5 T (MAGNETOM Sola, \linebreak[4]Siemens Healthcare, Erlangen, Germany), with an 18-channel body coil and a 32-channel spine coil (12 elements used). Three clinical T2w series of 2D thick slices are acquired in orthogonal orientations using an ultra-fast multi-slice HASTE sequence (TE=$90ms$, TR=$1200ms$, excitation/refocusing pulse flip angles of 90\textdegree/180\textdegree, interslice gap of 10\%, voxel size of $1.13 \times 1.13 \times 3.00 mm^3$). For consistency with the clinical fetal protocol, a limited field-of-view ($360 \times 360 mm^2$) centered on the above-referenced ROI is imaged. Each series contains 23 slices and is acquired in 28 seconds.

We extend the TE of this clinical protocol in order to acquire additional sets of three orthogonal series, leading to six configurations with 4, 5, 6, 8, 10, or 18 TEs uniformly sampled over the range of 90 ms to 298 ms. The acquisition time is about 90 seconds per TE, thus the total acquisition time ranges from 6 minutes (4 TEs) to 27 minutes (18 TEs). Binary masks are drawn on each LR series for reconstruction of a SR volume at every TE, as illustrated in Figure~\ref{fig1}-C.

\subsubsection{Gold-Standard Sequences for T2 Mapping.} A conventional single-echo spin echo (SE) sequence with variable TE is used as a reference for validation (TR=$5000ms$, 25 TEs sampled from 10 to $400ms$, voxel size of $0.98 \times 0.98 \times 6.00 mm^3$). One single 2D slice is imaged in 17.47 minutes for a given TE, which corresponds to a total acquisition time of more than 7 hours. As recommended by the phantom manufacturer, an alternative multi-echo spin echo (MESE) sequence is used for comparison purposes (TR=$5000ms$, 32 TEs equally sampled from 13 to $416 ms$, voxel size of $0.98 \times 0.98 \times 6.00 mm^3$). The total acquisition time to image the same 2D slice is of 16.05 minutes.

Gold-standard and HASTE acquisitions are made publicly available in our repository \cite{lajous_dataset_2020} for further reproducibility and validation studies.

\subsubsection{Evaluation Procedure.} We evaluate the accuracy of the proposed 3D SR T2 mapping framework with regard to T2 maps obtained from HASTE, MESE and SE acquisitions. Since only one single 2D coronal slice is imaged by SE and MESE sequences, quantitative measures are computed on the corresponding slice of the coronal 2D HASTE series and 3D SR images.
At a voxel-wise level, T2 standard deviation (SD) and R\textsuperscript{2} are computed to evaluate the fitting quality. A region-wise analysis is conducted over the three ROIs previously denoted as (a), (b) and (c). An automated segmentation of these areas in the HASTE and SR images is performed by Hough transform followed by a one-pixel dilation. Mean T2 values $\pm$ SD are estimated within each ROI.
The relative error in T2 estimation is computed using SE measurements as reference values. It is defined as the difference in T2 measures between either HASTE or SR and the corresponding SE reference value normalized by the SE reference value. This metric is used to evaluate the accuracy of the studied T2 mapping technique as well as its robustness to noise (see also Supplementary Material - Table I).

We run the same MRI protocol (SE, MESE and HASTE) on three different days in order to study the repeatability of T2 measurements. The relative error in T2 estimation between two independent experiments (i.e., on two different days) is calculated as described above, using every measure in turn as a reference. Thus, we are able to evaluate the mean absolute percentage error $|\Delta\varepsilon|$ as the average of relative absolute errors in T2 estimation over all possible reference days. The coefficient of variation (CV) for T2 quantification represents the variability (SD) relative to the mean fitted T2 value.

\section{Results}

\subsection{3D Super-Resolution T2 Mapping}
\begin{figure}[htb!]
\centering
\includegraphics[width=0.64\textwidth]{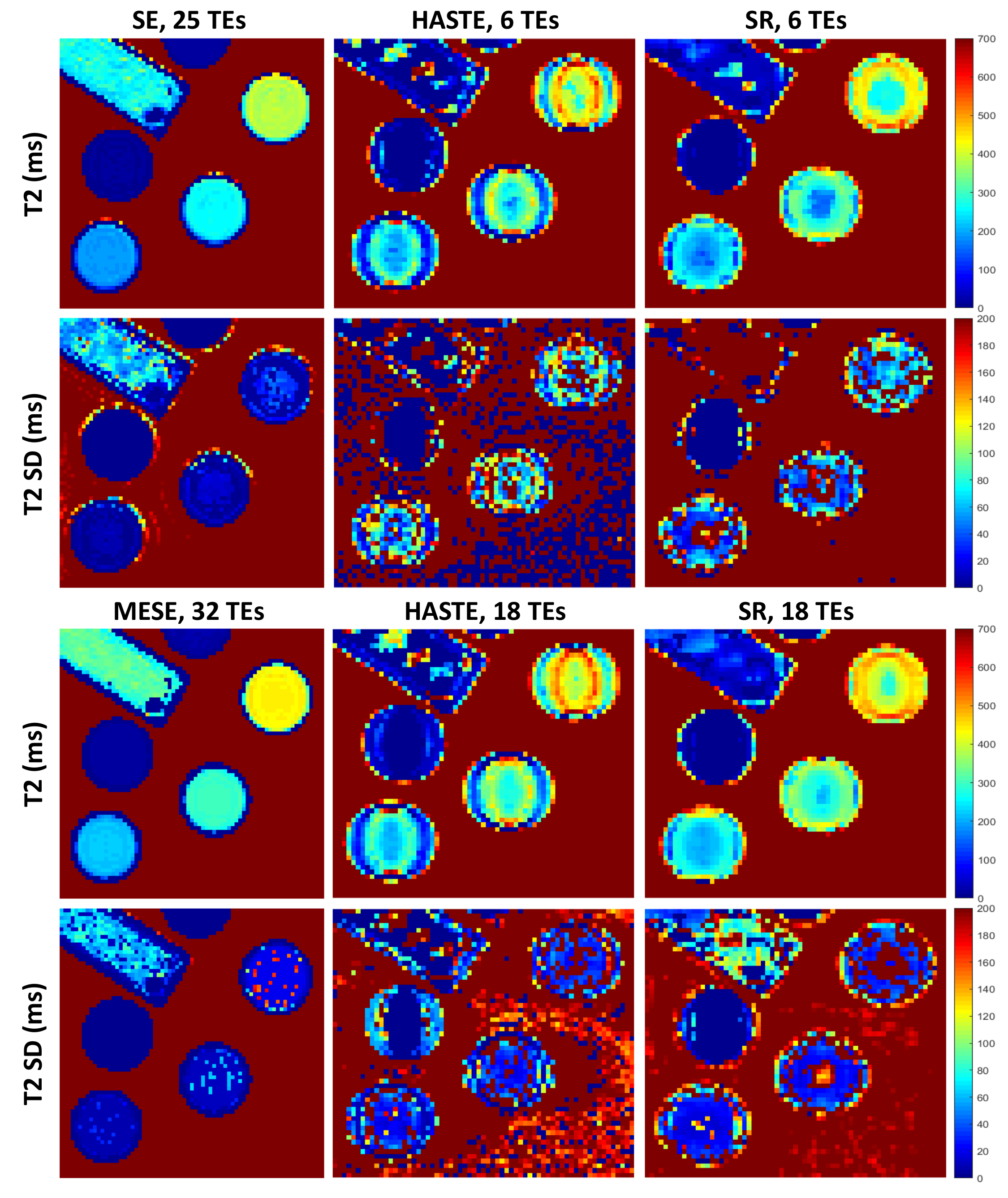}
\caption{Comparison of voxel-wise T2 maps and T2 SD maps estimated from SE, MESE, HASTE and corresponding SR reconstruction at variable TE}
\label{fig2}
\end{figure}

Voxel-wise T2 maps as derived from one coronal HASTE series and from the 3D SR reconstruction of three orthogonal HASTE series are shown in Figure~\ref{fig2} together with associated standard deviation maps. HASTE series show Gibbs ringing in the phase-encoding direction at the interface of the different elements. Since SE and MESE images are corrupted in a similar way across all TEs, a homogeneous T2 map is recovered for every element of the phantom (Figure~\ref{fig2}). Instead, as HASTE acquisitions rely on a variable k-space sampling for every TE, resulting T2 maps are subject to uncompensated Gibbs artifacts that cannot easily be corrected due to reconstruction of HASTE images by partial Fourier techniques~\cite{kellner_gibbs-ringing_2016}. Interestingly though, Gibbs ringing is much less pronounced in the SR reconstructions where it is probably attenuated by the combination of orthogonal series. Of note, \emph{in vivo} data are much less prone to this artifact.

\subsection{Repeatability Study}
As highlighted in Table~\ref{tab1}, T2 estimation is highly repeatable over independent measurements with a mean CV of less than 4\%, respectively 8\%, for T2 quantification from HASTE acquisitions, respectively SR with 5 TEs. The mean absolute percentage error is less than 5\% in HASTE images, respectively 10\% in SR.

\begin{table}
\caption{Repeatability of T2 mapping strategies between three independent experiments. Mean fitted T2 value $\pm$ SD, CV, mean absolute difference and mean absolute percentage error in T2 estimation are presented. The lowest difference for each (ROI, method) pair is shown in bold.}
\label{tab1}
\centering
\begin{tiny}
\begin{tabular}{|c|c|c|c|c|c|c|c|c|}
\hline
\multicolumn{2}{|c|}{} & SE & \multicolumn{6}{ c|}{HASTE/SR}\\
\hline
\multicolumn{2}{|c|}{TEs} & 25 & 4 & 5 & 6 & 8 & 10 & 18\\
\hline
\multirow{3}{*}{\rotatebox{90}{T2(ms)}}
& (a) & 380$\pm$7 & 422$\pm$2/288$\pm$7 & 427$\pm$12/312$\pm$21 & 425$\pm$2/352$\pm$13 & 451$\pm$4/409$\pm$18 & 454$\pm$0/403$\pm$20 & 451$\pm$2/407$\pm$12\\
& (b) & 256$\pm$5 & 314$\pm$6/188$\pm$8 & 304$\pm$12/223$\pm$11 & 315$\pm$4/258$\pm$5 & 337$\pm$6/287$\pm$8 & 335$\pm$9/288$\pm$4 & 333$\pm$10/297$\pm$4\\
& (c) & 187$\pm$2 & 252$\pm$9/159$\pm$10 & 247$\pm$2/180$\pm$14 & 249$\pm$4/207$\pm$6 & 267$\pm$1/225$\pm$5 & 267$\pm$6/223$\pm$4 & 263$\pm$3/233$\pm$4\\
\hline
\multirow{3}{*}{\rotatebox{90}{CV(\%)}} & (a) & 1.8 & 0.4/\textbf{2.3} & 2.8/6.7 & 0.5/3.8 & 1.0/4.4 & \textbf{0.0}/5.0 & 0.4/3.0\\
& (b) & 1.8 & 1.8/4.4 & 3.9/4.8 & \textbf{1.3}/2.0 & 1.8/2.9 & 2.8/1.6 & 2.9/\textbf{1.2}\\
& (c) & 1.0 & 3.6/6.2 & 0.8/7.7 & 1.5/2.9 & \textbf{0.3}/2.4 & 2.2/\textbf{1.7} & 1.1/1.9\\
\hline
\multirow{3}{*}{\rotatebox{90}{$|\Delta$T2$|$(ms)}} & (a) & 9.1 & 1.9/\textbf{8.9} & 15.8/24.4 & 2.7/17.8 & 5.2/23.3 & \textbf{0.2}/26.8 & 2.6/15.4\\
& (b) & 6.1 & 7.7/11.1 & 14.2/13.8 & \textbf{4.9}/6.0 & 7.7/10.8 & 11.4/5.6 & 12.8/\textbf{4.2}\\
& (c) & 2.1 & 12.1/11.8 & 2.6/16.4 & 4.5/7.7 & \textbf{1.1}/7.2 & 6.9/\textbf{4.9} & 3.8/5.3\\
\hline
\multirow{3}{*}{\rotatebox{90}{$|\Delta\varepsilon|(\%)$}} & (a) & 2.4 & 0.4/\textbf{3.1} & 3.7/7.7 & 0.6/5.1 & 1.2/5.7 & \textbf{0.0}/6.7 & 0.6/3.8\\
& (b) & 2.4 & 2.4/5.9 & 4.6/6.3 & \textbf{1.6}/2.3 & 2.3/3.8 & 3.4/1.9 & 3.9/\textbf{1.4}\\
& (c) & 1.1 & 4.8/7.3 & 1.1/9.4 & 1.8/3.7 & \textbf{0.4}/3.2 & 2.6/\textbf{2.2} & 1.4/2.3
\\
\hline
\end{tabular}
\end{tiny}
\end{table}

\subsection{Impact of the Number of Echo Times on T2 Measurements}
In an effort to optimize the acquisition scheme, especially regarding energy deposition and reasonable acquisition time in a context of fetal examination, we investigate the influence of the number of TEs on the T2 estimation accuracy. As T2 quantification is highly repeatable throughout independent measurements, the following results are derived from an arbitrarily-selected experiment.

T2 estimation by both clinical HASTE acquisitions and corresponding SR reconstructions demonstrates a high correlation with reference SE values over the 180-400 ms range of interest (Supplementary Material - Figure I).

\begin{figure}[htb!]
\centering
\includegraphics[width=0.8\textwidth]{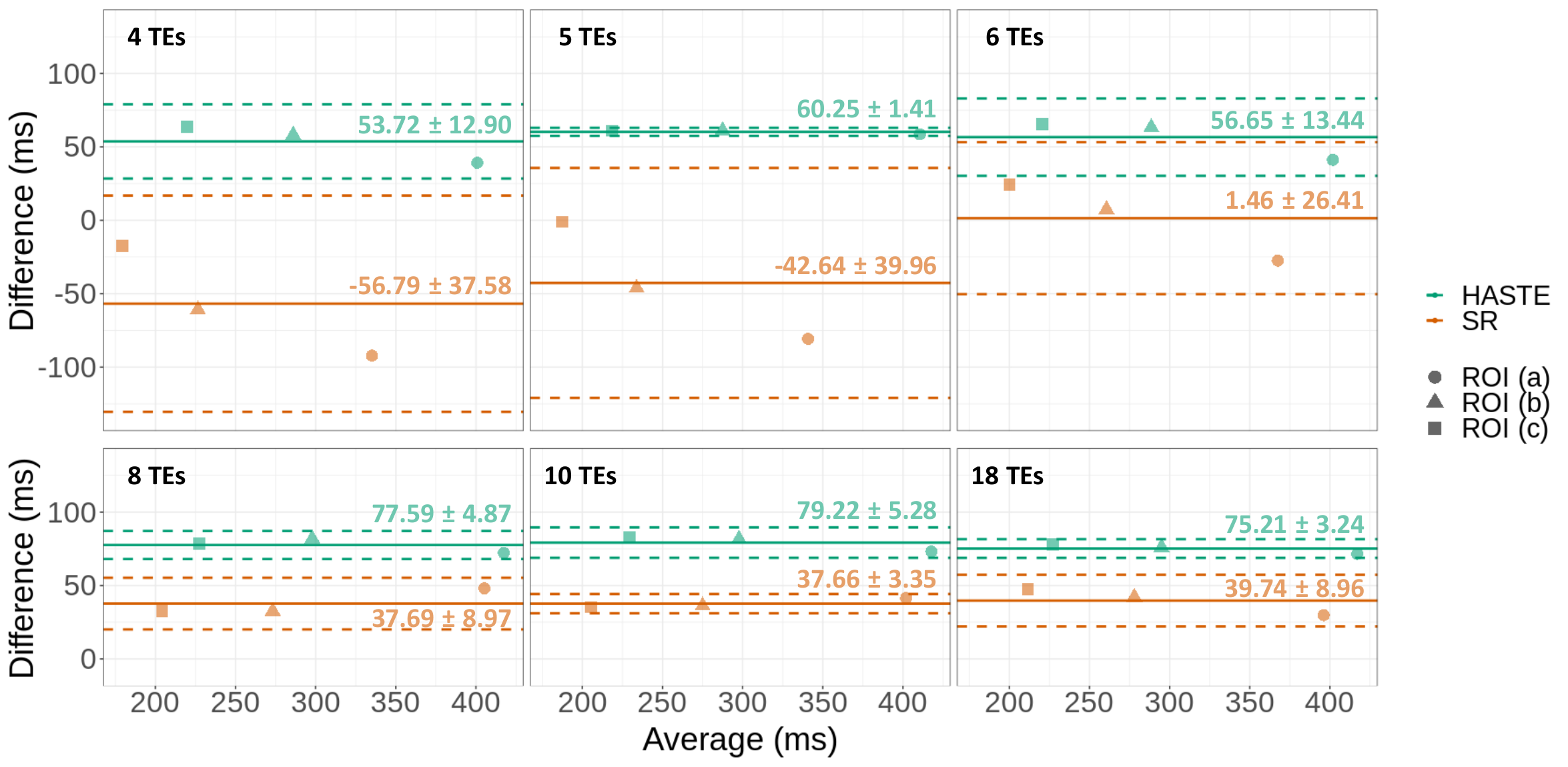}
\caption{Bland-Altman plots of differences in T2 quantification between HASTE / corresponding SR and reference SE in three ROIs for various numbers of TEs} \label{fig3}
\end{figure}

Bland-Altman plots presented in Figure~\ref{fig3} report the agreement between HASTE-/SR-based T2 quantification and SE reference values in the three ROIs. The average error in T2 estimation from HASTE series is almost the same across all configurations. The difference in T2 measurements is independent of the studied ROI. Conversely, the average error in SR T2 quantification varies with the number of TEs, the smallest average difference being for 6 TEs. In a given configuration, the difference in T2 measurements depends on the targeted value.

\begin{figure}[htb!]
\centering
\includegraphics[width=0.8\textwidth]{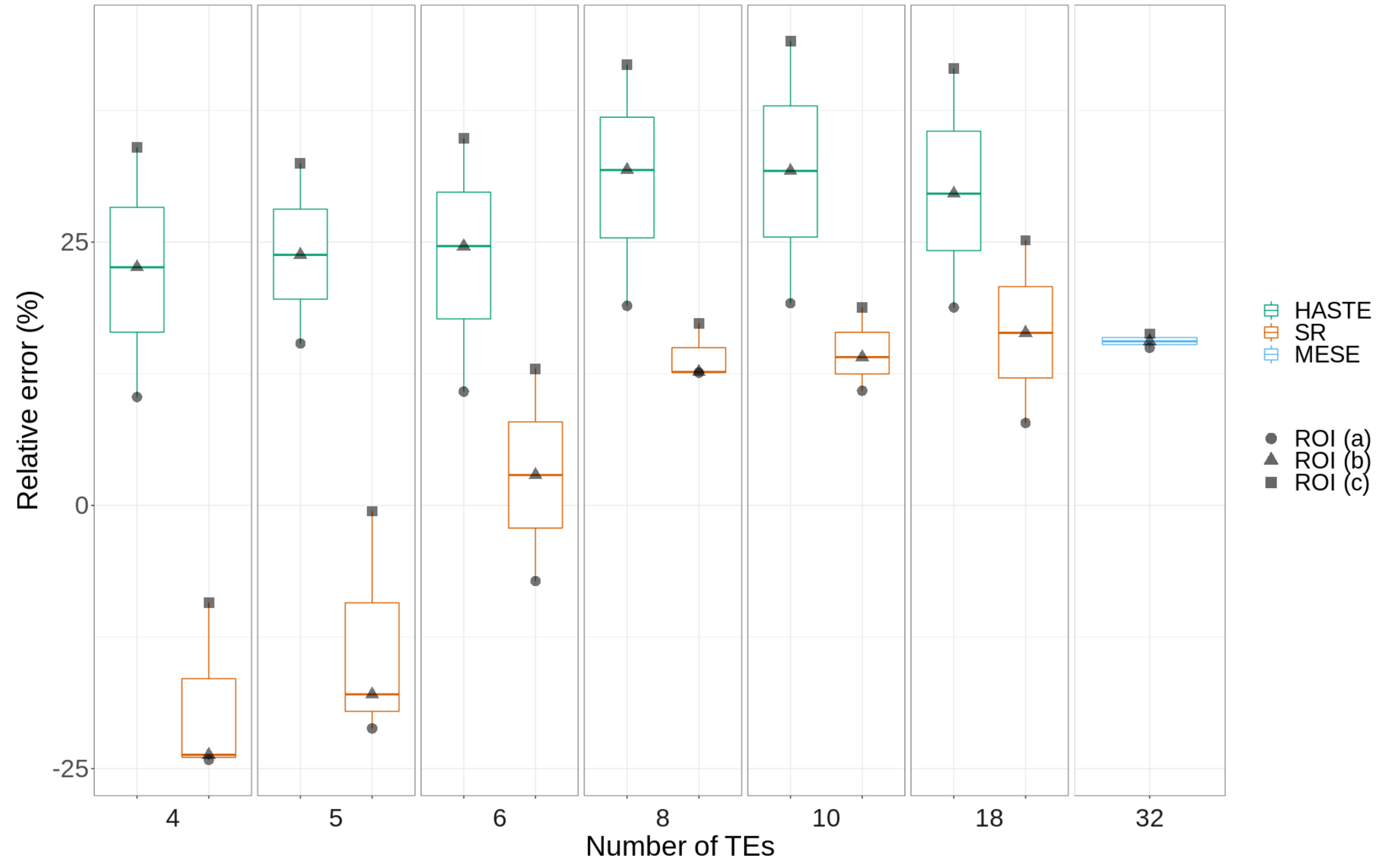}
\caption{Relative error in T2 quantification according to the method and number of TEs as compared to reference SE measurements}
\label{fig4}
\end{figure}

Figure~\ref{fig4} displays the relative error in T2 quantification from MESE, HASTE and SR as compared to SE measurements according to the number of TEs acquired. MESE provides T2 quantification with a small dispersion but with low out-of-plane resolution and prohibitive scanning times in a context of fetal MRI. In the following, its 16\%-average relative error is considered an acceptable reference error level. HASTE-based T2 quantification overestimates T2 values in the range of interest by around 25\%. As for MESE-based T2 mapping, such an overestimation can be attributed to stimulated echo contamination~\cite{mcphee_limitations_2018}. In the case of SR, quantification errors vary with the number of TEs acquired. Below six echoes, SR underestimates T2 values on average, and enables a dramatic improvement in T2 quantification over the HASTE-based technique, but only for T2 values less than 200 ms. Above 6 TEs, SR exhibits approximately the same average error as MESE. For six echoes, SR outperforms HASTE and MESE over the whole range of T2 values studied with a relative error less than 11\%.
Furthermore, preliminary results on T2 quantification from HASTE images corrupted by higher levels of noise and their corresponding SR reconstructions in this optimized set-up using six echoes demonstrate the robustness of the proposed SR T2 mapping technique (Supplementary Material - Table I).
Overall, SR substantially outperforms HASTE for T2 quantification.

\section{Conclusion}
This work demonstrates the feasibility of repeatable, accurate and robust 3D isotropic HR T2 mapping in a reasonable acquisition time based on SR-\linebreak[4]reconstructed clinical fast spin echo MR acquisitions. We show that SR-based T2 quantification performs accurately in the range of interest for fetal brain studies (180-430 ms) as compared to gold-standard methods based on SE or MESE. Moreover, it could be straightforwardly translated to the clinic since only the TE of the HASTE sequence routinely used in fetal exams needs to be adapted. A pilot study will be conducted on an adult brain to replicate these results \emph{in vivo}. Although our study focuses on static data, the robustness of SR techniques to motion makes us hypothesize that 3D HR T2 mapping of the fetal brain is feasible. The number of TEs required for an accurate T2 quantification in this context has to be explored.

\subsubsection{Acknowledgements.} This work was supported by the Swiss National Science Foundation through grants 141283 and 182602, the Center for Biomedical Imaging (CIBM) of the UNIL, UNIGE, HUG, CHUV and EPFL, the Leenaards and Jeantet Foundations, and the Swiss Heart Foundation. The authors would like to thank Yasser Alem\'an-G\'omez for his help in handling nifti images.


%
%
%
\bibliographystyle{splncs04}

\begin{thebibliography}{10}
\providecommand{\url}[1]{\texttt{#1}}
\providecommand{\urlprefix}{URL }
\providecommand{\doi}[1]{https://doi.org/#1}

\bibitem{abd_almajeed_myelin_2004}
Abd~Almajeed, A., Adamsbaum, C., Langevin, F.: Myelin characterization of fetal
  brain with mono-point estimated {T1}-maps. Magnetic Resonance Imaging
  \textbf{22}(4),  565--572 (2004). \doi{10/frdp45}

\bibitem{bano_model-based_2020}
Bano, W., Piredda, G.F., Davies, M., Marshall, I., Golbabaee, M., Meuli, R.,
  Kober, T., Thiran, J.P., Hilbert, T.: Model-based super-resolution
  reconstruction of {T2} maps. Magnetic Resonance in Medicine  \textbf{83}(3),
  906--919 (2020). \doi{10/gf85n4}

\bibitem{blazejewska_3d_2017}
Blazejewska, A.I., Seshamani, S., McKown, S.K., Caucutt, J.S., Dighe, M.,
  Gatenby, C., Studholme, C.: {3D} in utero quantification of {T2}* relaxation
  times in human fetal brain tissues for age optimized structural and
  functional {MRI}. Magnetic Resonance in Medicine  \textbf{78}(3),  909--916
  (2017). \doi{10/gf2n9z}

\bibitem{chen_t2_2018}
Chen, L.W., Wang, S.T., Huang, C.C., Tu, Y.F., Tsai, Y.S.: T2 relaxometry {MRI}
  predicts cerebral palsy in preterm infants. American Journal of
  Neuroradiology  \textbf{39}(3),  563--568 (2018). \doi{10/gdcz66}

\bibitem{deoni_quantitative_2010}
Deoni, S.C.: Quantitative relaxometry of the brain. Topics in Magnetic
  Resonance Imaging  \textbf{21}(2),  101--113 (2010). \doi{10/fj3m42}

\bibitem{dingwall_t2_2016}
Dingwall, N., Chalk, A., Martin, T.I., Scott, C.J., Semedo, C., Le, Q.,
  Orasanu, E., Cardoso, J.M., Melbourne, A., Marlow, N., Ourselin, S.: T2
  relaxometry in the extremely-preterm brain at adolescence. Magnetic Resonance
  Imaging  \textbf{34}(4),  508--514 (2016). \doi{10/ggb9qn}

\bibitem{ebner_automated_2020}
Ebner, M., Wang, G., Li, W., Aertsen, M., Patel, P.A., Aughwane, R., Melbourne,
  A., Doel, T., Dymarkowski, S., De~Coppi, P., David, A.L., Deprest, J.,
  Ourselin, S., Vercauteren, T.: An automated framework for localization,
  segmentation and super-resolution reconstruction of fetal brain {MRI}.
  NeuroImage  \textbf{206},  116324 (2020). \doi{10/ggdnsm}

\bibitem{gholipour_fetal_2014}
Gholipour, A., Estroff, J.A., Barnewolt, C.E., Robertson, R.L., Grant, P.E.,
  Gagoski, B., Warfield, S.K., Afacan, O., Connolly, S.A., Neil, J.J.,
  Wolfberg, A., Mulkern, R.V.: Fetal {MRI}: a technical update with educational
  aspirations. Concepts in Magnetic Resonance. Part A, Bridging Education and
  Research  \textbf{43}(6),  237--266 (2014). \doi{10/gf4bc6}

\bibitem{gholipour_robust_2010}
Gholipour, A., Estroff, J.A., Warfield, S.K.: Robust super-resolution volume
  reconstruction from slice acquisitions: application to fetal brain {MRI}.
  IEEE Transactions on Medical Imaging  \textbf{29}(10),  1739--1758 (2010).
  \doi{10/b2xmdp}

\bibitem{hagmann_t2_2009}
Hagmann, C.F., De~Vita, E., Bainbridge, A., Gunny, R., Kapetanakis, A.B.,
  Chong, W.K., Cady, E.B., Gadian, D.G., Robertson, N.J.: T2 at {MR} imaging is
  an objective quantitative measure of cerebral white matter signal intensity
  abnormality in preterm infants at term-equivalent age. Radiology
  \textbf{252}(1),  209--217 (2009). \doi{10/bqkd9r}

\bibitem{kainz_fast_2015}
Kainz, B., Steinberger, M., Wein, W., Kuklisova-Murgasova, M., Malamateniou,
  C., Keraudren, K., Torsney-Weir, T., Rutherford, M., Aljabar, P., Hajnal,
  J.V., Rueckert, D.: Fast volume reconstruction from motion corrupted stacks
  of {2D} slices. IEEE Transactions on Medical Imaging  \textbf{34}(9),
  1901--1913 (2015). \doi{10/f3svr5}

\bibitem{keenan_kathryn_e_multi-site_2016}
Keenan, K.E., {Stupic, Karl F}, {Boss, Michael A}, {Russek, Stephen E},
  {Chenevert, Tom L}, {Prasad, Pottumarthi V}, {Reddick, Wilburn E}, {Cecil,
  Kim M}, {Zheng, Jie}, {Hu, Peng}, {Jackson, Edward F}, {Ad Hoc Committee for
  Standards in Quantitative MR}: Multi-site, multi-vendor comparison of {T1}
  measurement using {ISMRM}/{NIST} system phantom. In: Proceedings of the 24th
  {Annual} {Meeting} of {ISMRM}. Singapore (2016), program number 3290

\bibitem{kellner_gibbs-ringing_2016}
Kellner, E., Dhital, B., Kiselev, V.G., Reisert, M.: Gibbs-ringing artifact
  removal based on local subvoxel-shifts. Magnetic Resonance in Medicine
  \textbf{76}(5),  1574--1581 (2016). \doi{10/f9f64r}

\bibitem{lajous_dataset_2020}
Lajous, H., Ledoux, J.B., Hilbert, T., van Heeswijk, R.B., Meritxell, B.C.:
  Dataset {T2} mapping from super-resolution-reconstructed clinical fast spin
  echo magnetic resonance acquisitions (2020). \doi{10.5281/zenodo.3931812}

\bibitem{leppert_t2_2009}
Leppert, I.R., Almli, C.R., McKinstry, R.C., Mulkern, R.V., Pierpaoli, C.,
  Rivkin, M.J., Pike, G.B., {Brain Development Cooperative Group}: T2
  relaxometry of normal pediatric brain development. Journal of Magnetic
  Resonance Imaging  \textbf{29}(2),  258--267 (2009). \doi{10/c77mvm}

\bibitem{mcphee_limitations_2018}
McPhee, K.C., Wilman, A.H.: Limitations of skipping echoes for exponential {T2}
  fitting. Journal of Magnetic Resonance Imaging  \textbf{48}(5),  1432--1440
  (2018). \doi{10/ggdj43}

\bibitem{milford_mono-exponential_2015}
Milford, D., Rosbach, N., Bendszus, M., Heiland, S.: Mono-exponential fitting
  in {T2}-relaxometry: relevance of offset and first echo. PLoS ONE
  \textbf{10},  e0145255 (2015). \doi{10/gfc68d}

\bibitem{nossin-manor_quantitative_2013}
Nossin-Manor, R., Card, D., Morris, D., Noormohamed, S., Shroff, M.M., Whyte,
  H.E., Taylor, M.J., Sled, J.G.: Quantitative {MRI} in the very preterm brain:
  assessing tissue organization and myelination using magnetization transfer,
  diffusion tensor and {T1} imaging. NeuroImage  \textbf{64},  505--516 (2013).
  \doi{10/f4jgtg}

\bibitem{rousseau_super-resolution_2010}
Rousseau, F., Kim, K., Studholme, C., Koob, M., Dietemann, J.L.: On
  super-resolution for fetal brain {MRI}. International Conference on Medical
  Image Computing and Computer-Assisted Intervention  \textbf{13}(Pt 2),
  355--362 (2010). \doi{10/bns47p}

\bibitem{schneider_evolution_2016}
Schneider, J., Kober, T., Bickle~Graz, M., Meuli, R., Hüppi, P.S., Hagmann,
  P., Truttmann, A.C.: Evolution of {T1} relaxation, {ADC}, and fractional
  anisotropy during early brain maturation: a serial imaging study on preterm
  infants. American Journal of NeuroRadiology  \textbf{37}(1),  155--162
  (2016). \doi{10/f7489d}

\bibitem{tourbier_sebastientourbiermialsuperresolutiontoolkit_2019}
Tourbier, S., Bresson, X., Hagmann, P., Meuli, R., Bach~Cuadra, M.:
  sebastientourbier/mialsuperresolutiontoolkit: {MIAL} {Super}-{Resolution}
  {Toolkit} v1.0 (2019). \doi{10.5281/zenodo.2598448}

\bibitem{tourbier_efficient_2015}
Tourbier, S., Bresson, X., Hagmann, P., Thiran, J.P., Meuli, R., Bach~Cuadra,
  M.: An efficient total variation algorithm for super-resolution in fetal
  brain {MRI} with adaptive regularization. NeuroImage  \textbf{118} (2015).
  \doi{10/f7p5zx}

\bibitem{travis_more_2019}
Travis, K.E., Castro, M.R.H., Berman, S., Dodson, C.K., Mezer, A.A.,
  Ben-Shachar, M., Feldman, H.M.: More than myelin: probing white matter
  differences in prematurity with quantitative {T1} and diffusion {MRI}.
  NeuroImage. Clinical  \textbf{22},  101756 (2019). \doi{10/ggnr3d}

\bibitem{vasylechko_t2_2015}
Vasylechko, S., Malamateniou, C., Nunes, R.G., Fox, M., Allsop, J., Rutherford,
  M., Rueckert, D., Hajnal, J.V.: T2* relaxometry of fetal brain at 1.5 {Tesla}
  using a motion tolerant method. Magnetic Resonance in Medicine
  \textbf{73}(5),  1795--1802 (2015). \doi{10/gf2pbh}

\bibitem{yarnykh_quantitative_2018}
Yarnykh, V.L., Prihod'ko, I.Y., Savelov, A.A., Korostyshevskaya, A.M.:
  Quantitative assessment of normal fetal brain myelination using fast
  macromolecular proton fraction mapping. American Journal of Neuroradiology
  \textbf{39}(7),  1341--1348 (2018). \doi{10/gdv9nf}

\end{thebibliography}

\end{document}


\begin{center}\fontsize{14pt}{17pt}\textbf{Supplementary material}\end{center}

\noindent\fontsize{12pt}{14pt}\textbf{Correlation of T2 quantification according to the number of echoes acquired}

\begin{figure}[htb!]
\centering
\includegraphics[width=0.8\textwidth]{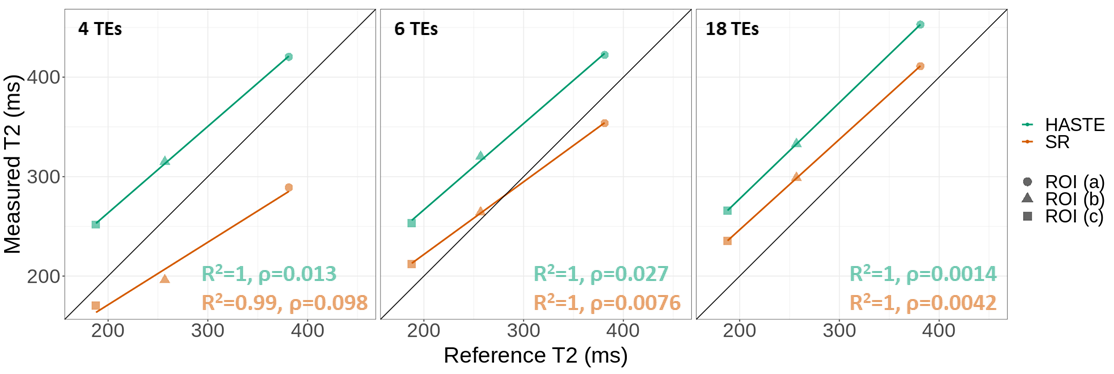}
\caption{Correlation plots of T2 quantification in three ROIs of the NIST phantom. HASTE- and SR-based T2 measurements from 4, 6 and 18 TEs respectively are compared to SE reference values (ground truth).}
\end{figure}

\newpage
\noindent\fontsize{12pt}{14pt}\textbf{Robustness of SR T2 mapping: a signal-to-noise ratio (SNR) study}

\begin{table}
\caption{Robustness to noise of SR T2 mapping. HASTE images of the NIST phantom acquired at 6 TEs were denoised using a non-local means algorithm. Various levels of Rician noise were added to the images before SR-reconstruction and voxel-wise T2 fitting. Reference data were obtained by the same post-processing to match the SNR of 370 in ROI (a) of the original MR acquisitions at TE=$90ms$. Mean T2 value $\pm$ SD, absolute difference and absolute percentage error in T2 estimation are presented. Results anticipate the robustness to noise of the proposed SR T2 mapping technique.}
\centering
\begin{footnotesize}
\begin{tabular}{|c|c|c|c|c|c|}
\hline
\multicolumn{2}{|c|}{} & \multicolumn{4}{ c|}{HASTE/SR}\\
\hline
\multicolumn{2}{|c|}{SNR in (a)} & 370 & 290 & 170 & 60\\
\hline
\multirow{3}{*}{\rotatebox{90}{T2(ms)}}
& (a) & 424$\pm$70/356$\pm$75 & 424$\pm$73/355$\pm$76 & 415$\pm$79/354$\pm$76 & 404$\pm$86/363$\pm$70\\
& (b) & 320$\pm$62/269$\pm$60 & 319$\pm$62/266$\pm$61 & 320$\pm$62/260$\pm$62 & 320$\pm$93/259$\pm$61\\
& (c) & 254$\pm$52/208$\pm$40 & 253$\pm$50/215$\pm$34 & 258$\pm$72/213$\pm$35 & 254$\pm$82/210$\pm$38\\
\hline
\multirow{3}{*}{\rotatebox{90}{$|\Delta$T2$|$(ms)}} & (a) & x & 0/1 & 8/2 & 19/7\\
& (b) & x & 1/4 & 0/10 & 0/10\\
& (c) & x & 0/7 & 4/5 & 1/2\\
\hline
\multirow{3}{*}{\rotatebox{90}{$|\Delta\varepsilon|(\%)$}} & (a) & x & 0.04/0.18 & 1.95/0.54 & 4.56/2.07\\
& (b) & x & 0.19/1.32 & 0.09/3.55 & 0.02/3.66\\
& (c) & x & 0.07/3.43 & 1.63/2.29 & 0.22/1.06\\
\hline
\end{tabular}
\end{footnotesize}
\end{table}